%% file: main.tex
\documentclass[letterpaper,twocolumn,10pt]{article}
\usepackage{usenix-2020-09}

\usepackage{tikz}
\usepackage{amsmath}

\input{macros}

\usepackage{url}

\usepackage{hyperref}
\usepackage{breakurl}
\usepackage{balance}
\usepackage{float}
\usepackage{tikz}
\usepackage{amsmath}
\usepackage{multirow}
\usepackage{booktabs}
\usepackage{xspace}
\usepackage{listings}
\usepackage{listings-rust}
\usepackage{color}
\usepackage{titling}
\usepackage{enumitem}
\usepackage[normalem]{ulem}

\begin{document}

\date{}

\title{\Large \bf SquirrelFS: using the Rust compiler to check file-system crash consistency}

\author{
{\rm Hayley LeBlanc}\\
University of Texas at Austin
\and
{\rm Nathan Taylor}\\
University of Texas at Austin
\and
{\rm James Bornholt}\\
University of Texas at Austin
\and
{\rm Vijay Chidambaram}\\
University of Texas at Austin
} 

\maketitle
\thispagestyle{empty}

\input{sections/abstract/abstract3}

\input{sections/introduction/introduction10}

\input{sections/background/background9}

\input{sections/combined/combined3}

\input{sections/experiences/experiences5}

\input{sections/evaluation/evaluation4}

\input{sections/related/related3}

\input{sections/conclusion1}


\section*{Acknowledgments}

\newcontent{We thank our anonymous shepherd, OSDI reviewers, and the members of SaSLab and LASR at UT Austin for their insightful comments and feedback. This work was supported by NSF CAREER \#1751277, NSF CCF \#2124044, and donations from Amazon, Toyota, and VMware.}

\bibliographystyle{plain}
\bibliography{refs.bib}

\end{document}

%% file: macros.tex
\newcommand{\sref}[1]{\S\ref{#1}}
\newcommand{\vheading}[1]{\vspace{0.05in}\noindent\textbf{#1}}

\newcommand{\myx}{$\times$\xspace}

\newcommand{\sysname}{{\textsc{SquirrelFS}}\xspace}

\newcounter{mylabelcounter}

\makeatletter
\newcommand{\labelText}[2]{%
#1\refstepcounter{mylabelcounter}%
\immediate\write\@auxout{%
  \string\newlabel{#2}{{1}{\thepage}{{\unexpanded{#1}}}{mylabelcounter.\number\value{mylabelcounter}}{}}%
}%
}
\makeatother

\newcounter{bugcounter}
\newcommand{\newcontent}[1]{#1}
\newcommand{\edited}[1]{#1}

\newcounter{observationcounter}

\newcounter{lessoncounter}

\usepackage{tikz}
\usepackage{pgfplots}
\usetikzlibrary{calc,trees,positioning,arrows,chains,shapes.geometric,%
  decorations.pathreplacing,decorations.pathmorphing,decorations.text,shapes,%
  matrix,shapes.symbols,patterns,shadows}

\RequirePackage{xcolor}
\definecolor{RED}{rgb}{1,0,0}
\definecolor{BLUE}{rgb}{0,0,1}
\definecolor{White}{rgb}{1,1,1}

\newcommand*\circled[1]{\tikz[baseline=(char.base)]{
            \node[shape=circle,fill,inner sep=0.75pt] (char) {\textcolor{white}{#1}};}}

%% file: sections/abstract/abstract3.tex
\begin{abstract}
  This work introduces a new approach to building crash-safe file systems for persistent memory.
  We exploit the fact that Rust's typestate pattern allows compile-time enforcement of a specific order of operations.
  We introduce a novel crash-consistency mechanism, \emph{Synchronous Soft Updates}, that boils down crash safety to enforcing ordering among updates to file-system metadata. 
  We employ this approach to build  \sysname, a new file system with crash-consistency guarantees that are \emph{checked at compile time}.
  \sysname avoids the need for separate proofs, instead incorporating correctness guarantees into the typestate itself.
  Compiling \sysname only takes tens of seconds; successful compilation indicates crash consistency,
   while an error provides a starting point for fixing the bug.
  We evaluate \sysname against state of the art file systems such as NOVA and WineFS, and find that \sysname achieves similar or better performance on a wide range of benchmarks and applications.
\end{abstract}

%% file: sections/introduction/introduction10.tex
\section{Introduction}
\label{sec:intro}

One of the most important properties for file systems is to preserve their integrity and user data in the face of a crash or a power loss \cite{ChidambaramPhd15, crash-cacm15, Bornholt16, McKusick99, hagmann87-cedar, Lorie77-ShadowPaging, HitzLM94}. 
Unfortunately, building crash-consistent file systems is challenging; checking or ensuring crash consistency is even more so~\cite{URLext4dataloss, chajed2022}.

There are two main approaches to building file systems today, as summarized in Table~\ref{tab:approach_comparison}.
First, we build file systems using low-level languages like C, and we use runtime testing to gain some confidence in the correctness of the systems \cite{Yang2004, Mohan18, Kim2019, Kalbfleisch22, LeBlanc23, Neal20}.
Note that this approach is necessarily incomplete; testing can only reveal bugs, not prove their absence. 
However, this approach allows rapid development, and entire testing ecosystems have sprung up around this basic approach, like the widely-used xfstests \cite{xfstests} and Linux Test Project \cite{ltp}.

A different approach to building file systems is to verify them: we write a high-level specification of correct behavior (including crash behavior) and then prove that the implementation matches the specification
~\cite{Chen15, Sigurbjarnarson16, Chen17, Hance20, chajed2022}.
This approach can prove that the implementation does not have certain classes of bugs;
however, it comes at a high cost. 
\edited{For each line of code in the implementation, we may need to write 7--13 lines of proof.}
Writing and maintaining proofs is time-consuming and requires specialized expertise, constraining rapid development.

In this work, we seek to find a middle ground between these two approaches. 
We would like to verify some aspects of file systems, but without the burden of having to write and maintain proofs. 
In particular, we are interested in \emph{crash consistency}, a correctness property that is especially difficult to test for.
In order to be crash consistent, systems must ensure that updates become persistent on storage media \emph{in the correct order}; however, hardware or caching layers may reorder updates to improve performance in unanticipated ways. 
Exposing crash-consistency bugs thus requires one to find and reproduce these low-level orderings, which requires specialized testing software \cite{Yang2004, Mohan18, Kim2019, Kalbfleisch22, LeBlanc23, Neal20}. 
Our goal is to develop lightweight approaches to statically check for crash-consistency bugs without the overhead of full verification.

\input{sections/figures/approach_comparison}

\input{sections/figures/typestate_overview}

We exploit two recent developments to achieve this goal (\sref{sec:background}).
\newcontent{First, the Rust programming language has a strong type system that supports powerful compile-time safety checks.
Our work takes inspiration from Corundum~\cite{Hoseinzadeh21}, a Rust crate (library) that uses Rust's type system to check low-level PM safety properties.
In this work, we observe that Rust's type system can also statically enforce that certain operations are carried out in a given order~\cite{Rust, gjenset}.}
Since the root of crash consistency is ordering updates to storage, if we can encode those ordering-based invariants in the type system,
the compiler can ensure the invariants hold at compile time. 

However, to do so, crash consistency must be derived purely from ordering-based invariants;
some mechanisms such as journaling use writes to a log to obtain atomicity, which is harder to encode in the type system.
Soft updates achieves crash consistency purely via ordering~\cite{McKusick99}, but the traditional soft updates scheme is complex and hard to implement~\cite{Frost07, Aurora09}. 

We observe that the low latency of persistent memory~\cite{Yang2004, Xu19pm} allows file-system operations to be \emph{synchronous}; all updates to storage media are durable by the time each operation returns~\cite{Xu16, Kadekodi19, Kadekodi21, Kwon17, Dullor14}.
We take advantage of persistent memory's synchronous updates and byte addressability
to develop a new mechanism for crash consistency we term \emph{Synchronous Soft Updates}. 

Synchronous Soft Updates builds on the classical soft updates mechanism~\cite{McKusick99}, but avoids most of the complexity that prevented the widespread adoption of soft updates~\cite{Aurora09}.
Two of the most complicated aspects of soft updates, dependency structures and cyclic dependency management, arise due to the need to track ordering requirements between block-sized updates across asynchronous operations.
Synchronous Soft Updates eliminates these challenges entirely by using fast, fine-grained storage to back synchronous operations.

We ensure that the ordering invariants of Synchronous Soft Updates hold by using the Rust compiler.
We take advantage of Rust's support for the \emph{typestate pattern}, an API design pattern where an object's type reflects the operations that have been performed on it \cite{Strom86}. 
The legal order of operations is encoded in function signatures and enforced by Rust's typechecker.
For example, an uninitialized inode has a different type than an initialized one; attempting to use one where the other is expected will result in a compile-time error.
Figure~\ref{fig:typestate_overview} illustrates the approach. 

We implement Synchronous Soft Updates in a new file system for PM  called \sysname and use the typestate pattern in Rust to check that update orderings are implemented correctly. \sysname provides crash-atomic metadata system calls, including {\tt rename}; on the original soft updates, a crash during rename could result in both the source and destination existing. 
\sysname compiles and typechecks in seconds, whereas running verification on existing storage systems takes minutes or hours. Building \sysname required no modifications to the Rust language.

We evaluate \sysname by comparing to a number of file systems meant for persistent memory, such as NOVA \cite{Xu16} and WineFS \cite{Kadekodi21} (\sref{sec:eval}). 
We use Intel's Optane DC Persistent Memory Module for our comparison, and find that \sysname offers comparable or better performance to other PM file systems across a broad range of workloads.
\newcontent{
The current \sysname prototype prioritizes simplicity of update ordering rules over performance in some areas, leading to relatively high mount times and memory utilization; however, these are not fundamental limitations of the design.
}
We also model the design of \sysname using the Alloy model-checking language \cite{alloybook} to gain confidence in the correctness of its Synchronous Soft Updates mechanism.

\newcontent{
We note that \sysname is not fully verified, and thus does not obtain the strong correctness guarantees of verified storage systems like FSCQ~\cite{Chen15}. 
Crash-consistency bugs may still occur in \sysname if their root causes are unrelated to ordering, if the ordering rules enforced by the compiler are incorrect, or if trusted code in \sysname's implementation or the Rust compiler are buggy.
For example, \sysname's ordering rules guarantee that inodes are always initialized before they are linked into the file system tree, but they do not guarantee that the contents of the inode are correct. 
\sysname's static checks are also limited by the capabilities of the Rust compiler.
For instance, the Rust compiler cannot check properties about variable-sized sets of data structures, as checking such properties is undecidable in general.
}

\sysname offers a useful new point in the spectrum of approaches to building \newcontent{robust} storage systems; 
it provides weaker guarantees than verified systems, but comes at a lower cost.
As such, we hope that it proves useful for developers of storage systems that require strong guarantees, good performance, and rapid development. 

In summary, this work makes the following contributions:
\begin{itemize}[noitemsep, parsep=0.1cm]
    \item Statically-checked crash consistency, an approach where high-level properties are encoded into the type system and checked at compile time (\sref{sec:sfs})
    \item The Synchronous Soft Updates crash-consistency mechanism for persistent-memory file systems (\sref{sec-ssu})
    \item The \sysname prototype, along with a discussion of lessons learned during its development (\sref{sec:lessons}).
\end{itemize}      

\sysname and its Alloy model are publicly available at \url{https://github.com/utsaslab/squirrelfs}.

%% file: sections/figures/approach_comparison.tex
\begin{table}[]
    \centering
    \begin{tabular}{l c c c}
        \toprule
         Approach & Complete & Dev effort & Time to check \\
         \midrule
         Testing & No & Low & Medium \\
         Verification & Yes & High & High \\
         This work & Yes & Medium & Low \\
        \bottomrule
    \end{tabular}
    \caption{Comparison of different approaches to ensuring crash consistency in file systems.}
    \label{tab:approach_comparison}
\end{table}

%% file: sections/figures/typestate_overview.tex
\begin{figure*}
    \centering
    \includegraphics[width=0.85\textwidth]{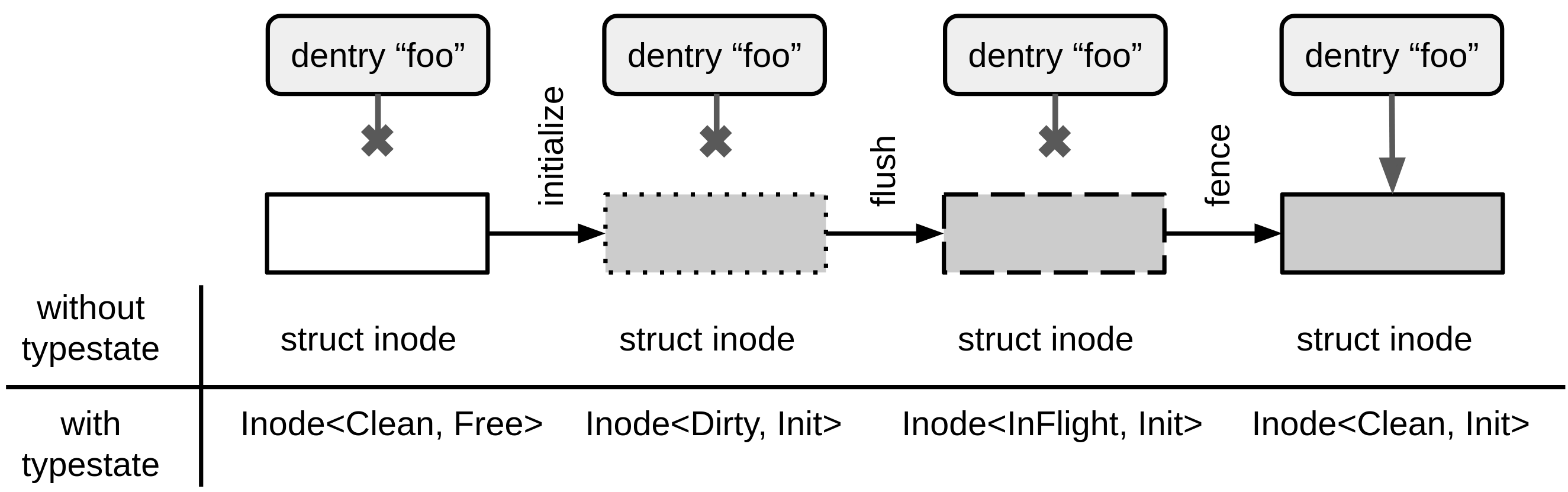}
    \caption{For a soft updates file system to be crash-consistent, directory entries should only point to fully-initialized, durable inodes. In existing file systems, all persistent inodes have the same type, regardless of whether they are durable or have been initialized. With typestate, durability and the inode's contents are reflected in its type.}
    \label{fig:typestate_overview}
\end{figure*}

%% file: sections/background/background9.tex
\section{Background and Motivation}
\label{sec:background}

\subsection{Crash Consistency}

A file system is \emph{crash consistent} if it can recover to a consistent state after a power loss or a crash~\cite{ChidambaramPhd15, crash-cacm15, Bornholt16}.
A consistent file system is one where all the metadata is in sync; for example, two files cannot (mistakenly) claim the same data block.
Files present before the crash must exist post-crash, and the data in files must remain valid.

\vheading{Crash-consistency mechanisms}. Crash consistency is generally achieved using mechanisms such as journaling~\cite{Needham88, hagmann87-cedar}, copy-on-write~\cite{btrfs, Lorie77-ShadowPaging, HitzLM94}, or soft updates~\cite{McKusick99}.
The root of crash consistency is correctly \emph{ordering} writes to storage \cite{ChidambaramPhd15}; for example, a data block must be initialized before a file points to it.
Soft updates achieves crash consistency by carefully ordering in-place updates to storage such that all possible crash states are consistent~\cite{McKusick99}.
To enforce ordering, soft updates must track updates across asynchronous operations and resolve cyclic dependencies when they arise.
Though soft updates is used in FreeBSD~\cite{McKusick14}, it has not been widely adopted due to its high complexity.

\vheading{Ensuring crash consistency}. Ensuring that a given file system achieves crash consistency is challenging.
There are two main approaches.
The first approach is testing, in which possible crash states of a file system are explored and checked for consistency.
Obtaining these crash states requires support from tools like eXplode~\cite{Yang2004}, CrashMonkey~\cite{Mohan18}, Hydra~\cite{Kim2019}, Chipmunk~\cite{LeBlanc23}, or Vinter~\cite{Kalbfleisch22}.
While such testing tools can find many bugs, they cannot prove overall correctness or the absence of crash-consistency bugs.

The second approach is to build verified file systems.
A developer writes a high-level specification of correctness and a lower-level implementation, and proves that the implementation satisfies the specification.
This approach is stronger than testing in that it can prove strong correctness properties and verify that there are no bugs.
\edited{However, it comes at a high cost: the developer has to write 7--13 lines of proof for every line of code.}
For example, BilbyFS~\cite{Amani16} required 13k lines of proof for 1k lines of implementation code; VeriBetrKV~\cite{Hance20} used 45K lines of proof for 6k lines of implementation.
Another verified file system, FSCQ~\cite{Chen15}, has interleaved proof and implementation code that is 10\myx the size of the most similar unverified system.

This heavy proof burden constrains development in a number of ways.
First, building a verified system requires proof-writing expertise, which restricts the set of developers who can work on it.
Second, proofs must be written in tandem with the code that they verify, which extends development time.
Finally, making changes to the system requires corresponding changes to the proofs, making maintenance slow and preventing rapid updates.

\newcontent{
Corundum~\cite{Hoseinzadeh21} is a Rust crate for PM systems that, like \sysname, uses the Rust type system to enforce certain low-level PM safety properties at compile time.
For example, Corundum ensures that every update to PM occurs in a logged transaction, and prevents the storage of pointers to volatile memory in durable structures.
\sysname was inspired by Corundum and aims to  enforce higher-level properties like file-system crash consistency with Rust.
}

\subsection{The opportunity: Rust and PM}

We observe an opportunity to ensure file-system crash consistency in a cheap manner.

First, we note that the Rust programming language can {\bf statically enforce a specific order on operations} via its support for the \emph{typestate pattern}~\cite{gjenset, rust-typestate}.
Briefly, the typestate pattern enables an object's runtime state to be encoded in its type~\cite{Strom86}. 
This state can be checked at compile time via typechecking, ensuring that a given operation can only occur on a specific type.
Typestate information is stored in zero-sized types that incur no runtime overhead.

For example, one consistency rule enforced by soft updates is that a directory entry should never point to an uninitialized inode.
Listing~\ref{lst:invalid_typestate} shows how typestate is used to enforce this rule.
To create a new file, we first \edited{obtain} a free directory entry and inode.
Initially, both objects have typestate {\tt Free}.
Then, we initialize the directory entry, transitioning its type to {\tt Dentry<Init>}.
The listing then has a bug in which the directory entry's inode number is set by {\tt commit\_dentry()} before the inode is initialized, breaking the consistency rule.
The Rust compiler catches this bug because the inode's current typestate {\tt Free} does not match the typestate {\tt Init} expected by the function.

\input{sections/figures/invalid_example}

Since soft updates is entirely built on ordering updates to file-system objects,
we can translate the required partial order into a set of types and use Rust's type checking to enforce the order. 
Thus, the invariants we want to maintain are translated into something the type system and compiler can enforce.
We note that we are able to do this with an \emph{unmodified} Rust compiler;
the new types introduced are no different to the compiler from existing types in the codebase. 

However, implementing soft updates correctly remains challenging even with typestate support.
With soft updates, file-system updates are applied to the page cache in DRAM, and then later written to storage in the right order.
Determining the right order requires tracking complex dependencies across asynchronous operations.
When a single file-system metadata object (such as an inode or a bitmap) is updated multiple times, it can lead to cyclic dependencies.

This leads to our second observation: persistent memory (PM) file systems support synchronous operations thanks to the low latency of the storage media~\cite{Yang20pm, Xu19pm}.
These file systems write updates directly to storage without first caching them in DRAM~\cite{Xu16, Kadekodi19, Kadekodi21, Kwon17, Dullor14}.
A {\bf synchronous implementation} of soft updates for persistent memory {\bf eliminates} the complexities of asynchronous dependency management, greatly simplifying the mechanism and allowing the relevant invariants to be encoded in Rust's type system.

%% file: sections/figures/invalid_example.tex
\begin{figure}
  \begin{lstlisting}[caption={
      The listing shows the typecheck process throwing an error when an uninitialized
      inode is passed to a function that expects an initialized inode.
},label=lst:invalid_typestate,captionpos=b,style=std]
fn new_file() {
    // Dentry<Free>
    let d = Dentry::get_free_dentry();  
    // Inode<Free>
    let i = Inode::get_free_inode();    
    // Dentry<Init>
    let d = d.set_name("foo");          
    let d = d.commit_dentry(i);
                  (* \color{red} \verb!^! expected `Inode<Init>`,*)
                    (* \color{red} found `Inode<Free>`*)
}
\end{lstlisting}
\end{figure}

%% file: sections/combined/combined3.tex
\section{SquirrelFS}
\label{sec:sfs}

We now present the design and implementation of \sysname,
a novel file system that uses the unmodified Rust compiler to check its crash consistency.
If the compilation is successful, it indicates that the ordering-based invariants
hold throughout the file system: in other words, the checking is complete. 
If compilation fails, the error reported by Rust is useful in figuring out which operations are out of order.
Compilation takes only seconds, offering quick feedback to developers.

\sysname is built on two key ideas:
\begin{itemize}[itemsep=0mm]
\item A novel crash-consistency mechanism, Synchronous Soft Updates, that achieves crash consistency purely via ordering file-system operations (\sref{sec-ssu})
  \item Using the Rust typestate pattern to encode ordering invariants into the Rust type system (\sref{sec-typestate})
\end{itemize}

It is important to note that we are not modifying the Rust compiler in any way.
To the Rust compiler, it is no different from type-checking any other code base;
we are merely using the type checking to ensure that crash consistency holds in the file system. 

We now describe the key ideas in more detail.

\subsection{Synchronous Soft Updates}
\label{sec-ssu}

We develop \emph{Synchronous Soft Updates} (SSU), a novel crash-consistency mechanism.
SSU is based on the traditional soft updates approach, but differs in two key aspects.
First, soft updates was designed for asynchronous settings, but all operations are synchronous in SSU.
Second, soft updates does not provide atomic rename; a crash during a rename of {\tt src} to {\tt dst} can result in both being present after a crash.
SSU fixes this flaw; renames are atomic, and a crash during rename will result in either {\tt src} or {\tt dst} after recovery. 

We now discuss why we developed SSU, its key aspects, and how renames are atomic in SSU. 

\vheading{Why a new mechanism?}
To go with our overall approach of encoding ordering-based invariants into the Rust type system,
we needed a mechanism that achieves crash consistency purely via ordering file-system updates.
This rules out mechanisms such as journaling and copy-on-write that use writes to a log or an extra copy to obtain atomicity. 
Soft updates~\cite{McKusick99} obtains crash consistency by enforcing ordering on in-place persistent updates to file-system objects; thus, it was a good match.
However, traditional soft updates suffered from two problems that we needed to tackle.
The first challenge was that soft updates had significant complexity arising from needing to track dependencies between asynchronous file-system operations;
the presence of cyclic dependencies also requires complex roll-back and roll-forward logic.
The second challenge is that soft updates does not provide atomic operations, particularly rename;
atomic rename is a crucial primitive for a number of POSIX applications~\cite{Pillai14}. 
Thus, we need to modify soft updates to tackle both its high complexity and lack of atomic operations.

\vheading{Synchronous operations}.
We observe that the root of complexity in soft updates (such as cyclic dependencies and structures for tracking dependencies) is \emph{asynchrony}. 
A \emph{synchronous} implementation of soft updates neatly avoids these complexities.
All updates would be made durable by the end of each system call, which would eliminate the need to track cross-operation dependencies.
Cyclic dependencies would no longer arise because there are no pending updates that can conflict with each other.
The SoupFS \cite{Dong17} soft updates file system for persistent memory eliminated cyclic dependencies using fine-grained updates,
but still required asynchronous dependency tracking.
A synchronous implementation is necessary to overcome both sources of complexity.

A synchronous version of soft updates was not feasible until now, as running this on magnetic hard drives or even solid state drives would be prohibitively slow.
However, synchronous soft updates is a good match for persistent memory (PM) due to its low latency;
system calls in many existing PM file systems are already synchronous~\cite{Xu16, Kadekodi19, Kadekodi21, Dullor14}. 

\newcontent{
Similar to traditional soft updates, SSU maintains crash consistency by enforcing ordering among updates to file-system objects.
SSU implements the original soft updates rules~\cite{Ganger94}:
\begin{enumerate}
    \item Never point to a structure before it has been initialized;
    \item Never re-use a resource before nullifying all previous pointers to it;
    \item Never reset the old pointer to a live resource before the new pointer has been set.
\end{enumerate}
These rules are significantly easier to enforce in a synchronous setting, as there is no need to track dependencies across asynchronous operations.
Like soft updates, SSU focuses on the integrity of file system metadata and cannot guarantee that operations on file data are atomic. 
SSU could be combined with journaling or copy-on-write to obtain stronger data guarantees.
}

\vheading{Atomic rename in SSU}.
SSU ensures renames are atomic by cleaning up file-system state after a crash.
In traditional soft updates, if there is a rename from {\tt src} to {\tt dst}, it is impossible to tell after a crash whether {\tt src} or {\tt dst} should be removed. 
To resolve this, SSU adds an extra field, called the \emph{rename pointer}, to directory entries in order to persistently save enough information to complete the rename operation after a crash. 
The rename pointer in the destination directory entry points to the physical location of the source directory entry.
\newcontent{
The rename pointer allows the file system to follow soft updates rule 3 (never reset the old pointer before the new one has been set) while also retaining the ability to distinguish between {\tt src} and {\tt dst} after a crash.
}

Note that this is similar to what journaling-based file systems do;
they write a log entry specifying {\tt src} and {\tt dst} so that the right clean-up action can be performed.
In SSU, the information in this log entry is distributed over the source and destination inodes;
taken together, they provide enough information to the file system.

Figure~\ref{fig:rename} illustrates the process. 
Step \circled{1} shows an example system state prior to the {\tt rename} operation.
In \circled{2}, {\tt dst}'s rename pointer (dotted line) is set to {\tt src}.
{\tt dst} is invalid, and {\tt src} is still valid.
In \circled{3}, we make {\tt dst} valid; this also logically invalidates {\tt src}.
This is an atomic point; after this step, the file system will always complete the rename operation.
If the file system crashes prior to this step, the rename pointer is cleared on recovery.
In \circled{4}, we physically mark {\tt src} as invalid.
In \circled{5}, the rename pointer is cleared, and in \circled{6} {\tt src} is fully deallocated.
Each step either modifies metadata that is invisible to the user (e.g., deallocating an orphaned directory entry) or atomically modifies a single 8-byte value.
All modifications must be durable before proceeding to the next step.

\input{sections/figures/rename_diagram}

A question that arises is how the file system finds {\tt src} and {\tt dst}. 
This is an example of how SSU is tailored for PM file systems.
In PM file systems, it is common for the file system to scan persistent objects to construct indexes in DRAM;
we add the rename-recovery procedure into this scan.
Thus, when building volatile indexes after a crash, the file system also looks for and completes any partially completed rename operations. 

\subsection{Using Rust to enforce ordering}
\label{sec-typestate}

\input{sections/figures/extended_typestate_listing}

Rust's typestate pattern can be used to ensure that \emph{a set of functions are always called in certain partial order}.
A total order is not necessary, as many operations involve independent updates that can be safely reordered.
As we discussed previously (\sref{sec:background}), an object's typestate is encoded in generic type parameters in its definition, and the partial order is encoded in the function signatures of its associated functions.

We encode two states (as different type parameters) in the types of persistent objects:
\begin{itemize}
\edited{\item \emph{Persistence} typestate is a representation of whether an object's most recent update(s) have been made durable. \edited{We use three persistence typestates: {\tt Dirty}, {\tt InFlight}, and {\tt Clean}.}
\item \emph{Operational} typestate represents the operations that have been performed on an object and is used to determine what operations can happen next.}
\end{itemize}

\edited{Persistence and operational} typestate are separate to capture the fact that most storage devices do not synchronously flush updates.
For example, in persistent memory, updates go to the CPU caches first, and must be explicitly flushed to the persistent media.

Listing~\ref{lst:ext_typestate} shows implementations of several methods of persistent {\tt Inode} and {\tt Dentry} types with \edited{persistence and operational} typestate as generic type parameters.
\edited{The methods {\tt flush} and {\tt fence} invoke a cache line write back and store fence respectively and are generic with respect to operational typestate.}
These methods must be used to ensure updates are persistent before continuing; 
for example, {\tt commit\_dentry()} requires an {\tt Inode<Clean, Init>} to ensure the inode's initialization cannot be transparently reordered with the directory entry updates.

This formulation of persistence typestate has several performance benefits.
First, because the {\tt flush} and {\tt fence} methods can only be called on an object whose typestate indicates it is not yet persistent, typechecking will prevent redundant persistence operations (thereby improving performance).
Second, developers can wait to flush updates until it is strictly necessary and can write additional transitions to enable multiple updates to share a single fence.

\vheading{Why Rust?} In order to obtain useful compiler-checked guarantees from the typestate pattern, each object must have exactly one typestate~\cite{Strom86}. Thus, languages with unrestricted aliasing (e.g., C) cannot support the typestate pattern, as different aliases for the same value can have different types.
Rust supports the pattern via its ownership type system, which ensures that each value has exactly one owner (and thus exactly one type).

\input{sections/figures/mkdir_diagram}

\subsection{Example: {\tt mkdir}}
\label{sec-example}

We use {\tt mkdir} to illustrate the typestates and dependency rules used in SSU.
To be crash consistent, an SSU implementation of {\tt mkdir} must ensure
(1) that a structure never points to an uninitialized resource, and (2) that each inode's link count is greater than or equal to its actual number of links. 
Both rules prevent dangling links in the event of a crash.

Figure~\ref{fig:mkdir} illustrates the dependencies in a {\tt mkdir} operation.
During {\tt mkdir}, three file-system objects are modified: an inode for the new directory, a directory entry for the new directory, and the inode of the parent directory.
Note that all three can be modified at the same time in a concurrent fashion, and can share a single store fence at the end (not shown).
\sysname uses volatile allocation structures, so they are not persisted during {\tt mkdir}.

The system first finds the parent inode and obtains a free directory entry in one of the parent's pages as well as a free inode. 
The inode is then initialized (i.e., setting its inode number, link count, timestamps), the directory entry's name is set, and the parent inode's link count is incremented. 

Next, we commit the directory entry by setting its inode number. 
This makes the directory entry valid and connects the inode to the file system tree. 
Directory entry commit is dependent on inode and directory entry initialization and parent link increment. 
Committing the directory entry before initializing the inode can result in a directory entry pointing to a garbage inode;
committing before incrementing the parent's link count can lead to dangling links.

\subsection{Implementation}
\label{sec-impl}

\input{sections/figures/squirrelfs_overview}

We implemented \sysname in Rust with 7500 LOC.
It uses bindings from the Rust for Linux project~\cite{RustForLinux} to connect to the Linux Virtual File System (VFS) layer. 
Figure~\ref{fig:squirrelfs_overview} shows \sysname's architecture.
We also built a model of \sysname in the model-checking language Alloy \cite{alloybook} to check its design for crash consistency issues.
We describe our experience developing \sysname in \sref{sec:lessons}.

\vheading{Overview}.
The design of \sysname combines aspects of FreeBSD's FFS~\cite{McKusick14} and PM file systems such as NOVA~\cite{Xu16} and WineFS~\cite{Kadekodi19}.
Like FFS, it has a simple on-storage layout, and uses soft updates.
Like other PM file systems, \sysname uses volatile index structures that are built when the file system is mounted.

\sysname's design was primarily influenced by two factors.
First, we wanted to keep dependencies as simple as possible and avoid nested persistent structures that are difficult to represent in typestate.
\newcontent{
Second, we assume the x86 PM persistence model in which only aligned updates of 8 bytes (or smaller) are crash atomic~\cite{Dullor14}. 
Under the x86 model, persistent addresses can be accessed via regular memory stores, but the corresponding cache line must be flushed before updates become persistent; a memory barrier like a store fence must also be invoked to correctly order stores~\cite{Rudoff17}.
Durable structures may also be updated via cache-bypassing non-temporal store instructions, which still require a store fence for persistence ordering.
This programming model influences the structure of persistent objects and restricts the set of legal orderings.
}

\newcontent{
All system calls in \sysname are synchronous, meaning that updates to durable structures made by each system call are durable by the time the system call returns. 
As such, {\tt fsync} is a no-op in \sysname.
Metadata-related operations are also crash-atomic.
Data-related operations are not atomic in the current \sysname prototype, which matches the default behavior of other PM file systems like NOVA~\cite{Xu16}. 
These operations could be made atomic by using copy-on-write to update file contents.
}

\vheading{Persistent layout}.
\sysname uses a simple layout to reduce the complexity of update dependencies.
\sysname splits the storage device into four sections: the superblock, the inode table, the page descriptor table, and the data pages.
The inode table is an array of all of the inodes in the system.
\sysname reserves enough space for approximately one inode for every 16KB of data (four pages), the same ratio used by the Linux Ext4 file system.

The page descriptor array contains page metadata.
Rather than having inodes point to the pages they own, each page descriptor contains a backpointer to its owner (similar to NoFS~\cite{Chidambaram12}) and stores its own metadata (e.g., its offset in the file).
This approach simplifies dependency rules for updates involving page allocation and deallocation.
All remaining space after the page descriptor table is used for data and/or directory pages.

\vheading{Volatile structures}.
\sysname's persistent layout simplifies typestate and update dependency rules, but it is not amenable to fast lookups. 
Therefore, \sysname uses indexes in DRAM to speed up lookup and read operations.
Each inode in the VFS inode cache has a private index for the resources it owns; index data for uncached nodes is stored in the VFS superblock.

Like many other PM file systems, \sysname uses volatile allocators: allocation information is not stored in a persistent manner, but rather rebuilt each time the file system is mounted.
It uses a per-CPU page allocator and a single shared inode allocator (which could be converted to a per-CPU allocator to improve scalability). 
The allocators use free lists backed by kernel RB-trees.

\sysname's indexes and allocators are rebuilt by scanning the file system when \sysname is mounted. 
An inode, directory entry, or page descriptor is considered allocated if \emph{any} of its bytes are non-zero. 
Directory entries and page descriptors are only valid if their inode numbers are set; inodes are valid only if they are reachable from the root.
Thus, updates that allocate new structures and set non-inode metadata fields need not be crash-atomic.

\newcontent{
\vheading{Synchronous Soft Updates.} \sysname uses an implementation of SSU for crash consistency. 
As shown in Figure~\ref{fig:mkdir}, operations that involve creation of new objects must first durably allocate and initialize resources before linking them into the file system (setting the directory entry's inode in the example) to enforce rule 1 (never point to a structure before it has been initialized). 
Deallocation proceeds in reverse; links are first cleared, then the object itself is deallocated by zeroing all of its bytes.
\sysname enforces rule 2 of soft updates (never re-use a resource before nullifying all previous pointers to it) by treating durable objects that are not completely zeroed out as allocated and by ensuring via typestate that pointers to the object are cleared before the object can be zeroed.
}

\newcontent{
\vheading{Typestate transition functions.} \sysname updates the typestate of objects via \emph{typestate transition functions}. 
These functions take ownership of the original object, modify it, and return it to the caller with the new typestate.
These functions are defined only on certain typestates to ensure they are called in a safe order.
For example, the typestate transition function {\tt commit\_dentry()}, shown in Listing~\ref{lst:ext_typestate}, is only defined for directory entries with type {\tt Dentry<Clean, Alloc>}, and also takes ownership of an inode of type {\tt Inode<Clean, Init>}.
Calling {\tt commit\_dentry()} out of order -- e.g., on a directory entry that has not yet been persistently allocated -- is a potential crash-consistency bug and results in a compiler error.
}

\newcontent{
\vheading{Concurrency.} \sysname supports concurrent file-system operations. 
It relies on VFS-level locking on durable resources like inodes.
This locking, together with Rust's type system, ensures that each resource has only one owner -- and only one type -- at any time, enabling strong typestate-based compile-time checking.
\sysname uses internal locks to protect its allocators and indexes.
}

\vheading{Building a model with Alloy}. 
While the typestate pattern can enforce a given operation order, it cannot verify that this order is crash consistent.
To gain more confidence that \sysname's design is crash consistent, we built a model of \sysname in the Alloy model checking language~\cite{alloybook}.

Alloy provides a language for specifying transition systems and a model checker to explore possible sequences of states (traces) of these systems.
Alloy's implementation is based on a logic of relations; each system is composed of a set of constraints that define a set of structures and the relations between them, and the model checker uses constraint solving to find traces.

In \sysname, there is roughly a one-to-one mapping between typestate transitions in the Rust implementation and the next-state predicates in the Alloy model.
Each next-state predicate specifies the states in which the transition may occur and the changes it makes to the model's state.
The model includes next-state predicates for typestate transitions and persistent updates. 
It also includes transitions that model crashes and recovery, which let us check \sysname's design for crash-consistency bugs.

Each persistent structure in \sysname is represented by a corresponding structure, also called a signature, in Alloy. 
The model also includes a {\tt Volatile} signature that is used to model volatile aspects of the file system like its indexes.
Each typestate is represented by a signature, and instances of persistent structures are mapped to their current typestate. 
Each file system operation is also represented by a signature, and relations map system calls to instances of persistent objects they are operating on as well as other volatile state (e.g., the locks held by that operation).
We use this to model concurrent file-system operations.

\subsection{Limitations of the approach}
\label{sec-limitations}

It is important to note that the typestate-based approach used in \sysname is not as powerful as full verification.
\edited{Fully-verified systems, such as the FSCQ file system~\cite{Chen15}, use theorem provers that can prove a wide variety of complex properties.
For example, a developer could prove, if required, that the system only uses even-numbered inodes for files.}

In contrast, our typestate-based approach can only check \emph{ordering-based} invariants.
Our approach could be used to verify that functions are called in a specific order;
for example, our approach can ensure that a file is not linked into the file-system tree before it is allocated.
However, it does not verify the implementation of each function that is called.

Thus, full verification is significantly more powerful and general, but it pays a cost in terms of complexity and development time.
Our approach is more targeted and ordering-based, but allows quick feedback and incremental development.

We believe this approach is a valuable addition to the repertoire of tools we have for building correct file systems.
This approach should be used alongside runtime testing and model-checking approaches. 

\subsection{Relevance beyond PM}
\label{sec-pm}

While we have designed \sysname for persistent memory,
\sysname would be relevant for any storage technology with byte-addressability and low latency.
The Compute Express Link \cite{cxl_spec} standard will support attached memory devices, including PM, via the Type 3 (CXL.mem) protocol. 
These CXL-attached PM devices will have the same interface and persistence semantics as current NVDIMMs, though performance will be lower~\cite{balcer}.

\sysname, and SSU file systems in general, could be used on CXL-attached memory. 
As \sysname's mount performance and memory footprint are tied to the size of the device, they may worsen with significantly larger-capacity devices.
Further work will be required to optimize file systems based on our approach for such devices.

%% file: sections/figures/rename_diagram.tex
\begin{figure}
    \centering
    \includegraphics[width=\columnwidth]{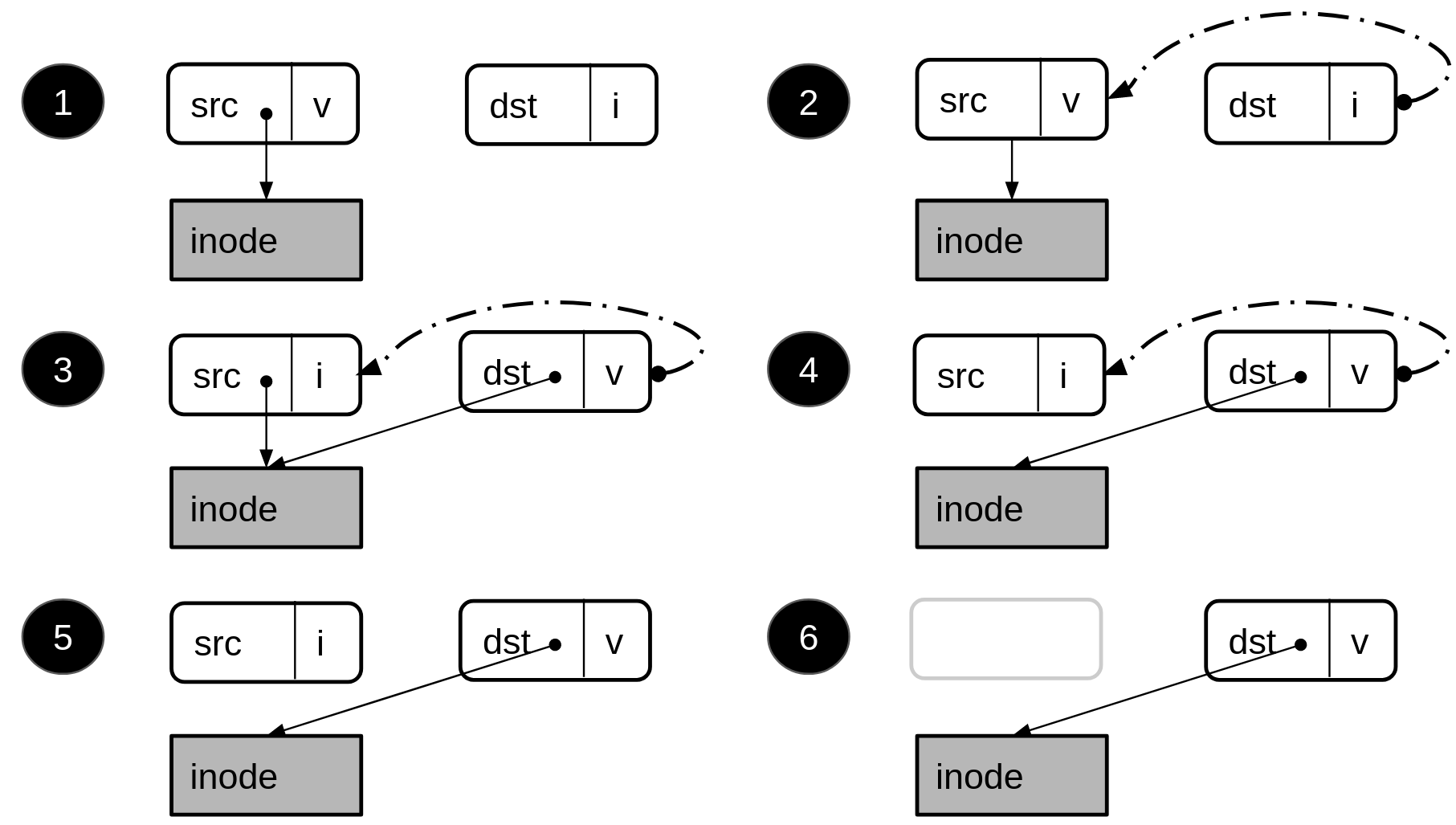}
    \caption{The figure shows the steps in atomic soft updates {\tt rename}. The dotted lines represent rename pointers and the solid lines represent inode pointers. {\tt src} and {\tt dst} are directory entries. The labels "v" and "i" indicate whether a directory entry is valid or invalid.}
    \label{fig:rename}
\end{figure}

%% file: sections/figures/extended_typestate_listing.tex
\begin{lstlisting}[language=Rust,float,style=boldts,caption={
Pseudocode implementations of file system objects with persistence and operational typestate. Typestate arguments are shown in bold.
},label=lst:ext_typestate,captionpos=b]
impl Inode<Clean, Free> {
    fn init_inode(self, ino: u64, ...) 
        -> Inode<Dirty, Init> {...}
}
impl Dentry<Clean, Alloc> {
    fn commit_dentry(
        self, 
        inode: Inode<Clean, Init>
    ) -> Dentry<Dirty, Committed> {...}
}
impl<S> Inode<Dirty, S> {
    fn flush(self) ->
        Inode<InFlight, S> {...}
}
impl<S> Inode<InFlight, S> {
    fn fence(self) -> 
        Inode<Clean, S> {...} 
} 

\end{lstlisting}

%% file: sections/figures/mkdir_diagram.tex
\begin{figure}[t]
    \centering
    \includegraphics[width=0.95\columnwidth]{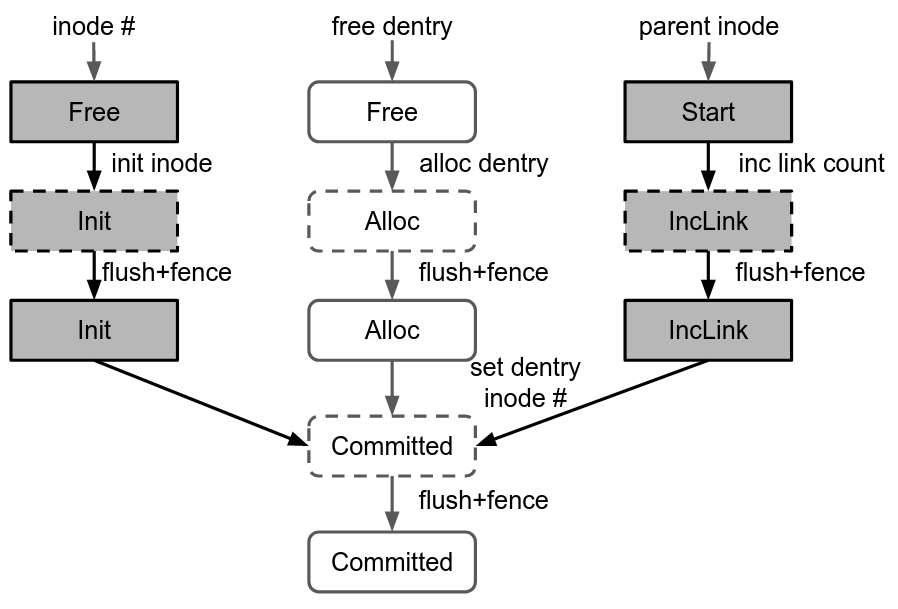}
    \caption{The figure shows the persistent updates and corresponding dependencies made during {\tt mkdir}. Inodes are dark gray and directory entries are white. Each object is labeled with its operational typestate and its outline indicates whether it is clean (solid) or dirty (dotted).}
    \label{fig:mkdir}
\end{figure}

%% file: sections/figures/squirrelfs_overview.tex
\begin{figure}[t]
    \centering
    \includegraphics[width=\columnwidth]{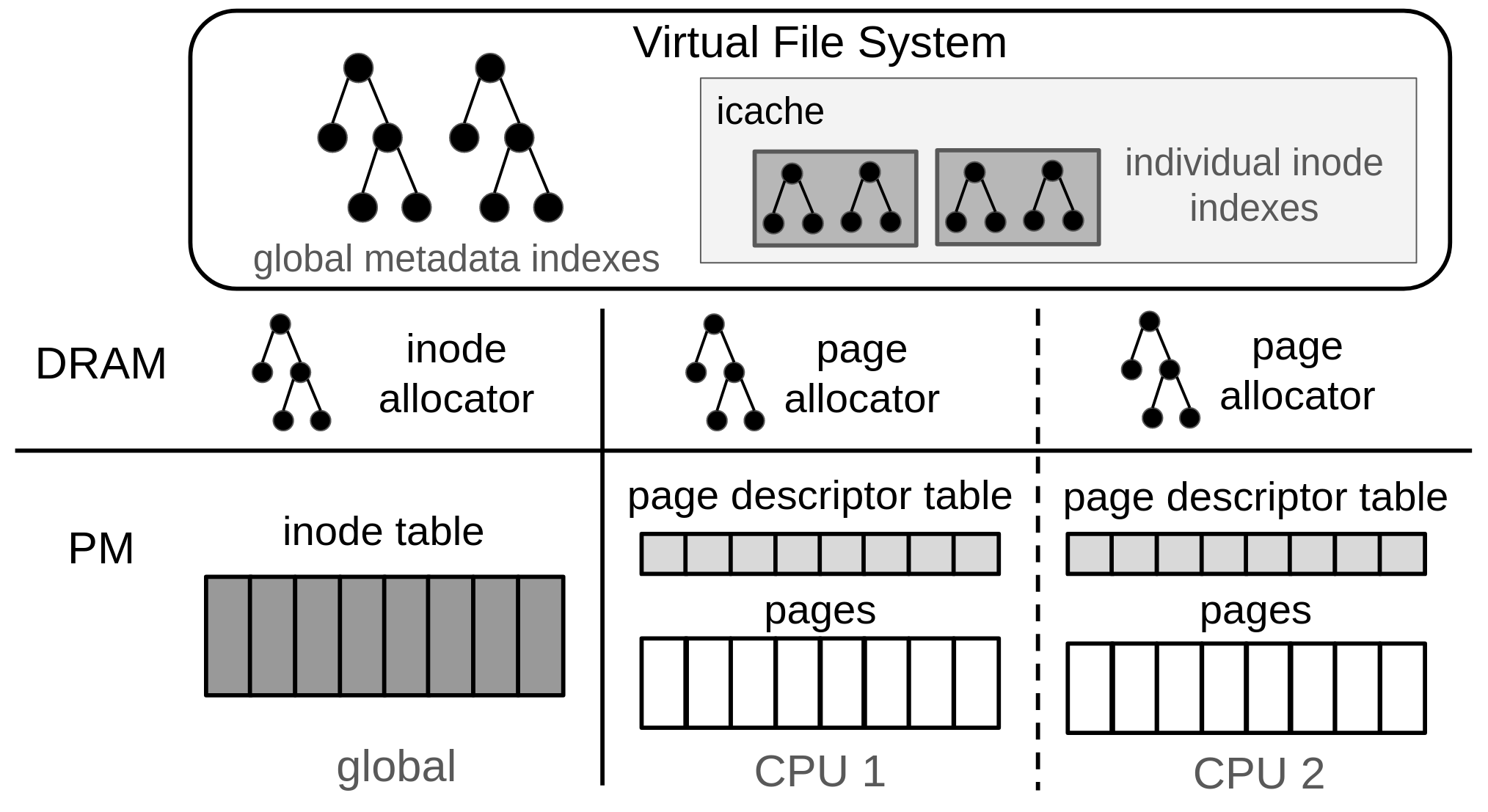}
    \caption{The figure shows the main components of \sysname. Each CPU has its own pool of pages and private page allocator. The inode allocator is shared between all CPUs. Volatile indexes are stored in VFS data structures.}
    \label{fig:squirrelfs_overview}
\end{figure}

%% file: sections/experiences/experiences5.tex
\section{Experience developing \sysname}
\label{sec:lessons}

We now describe our experience with designing, developing, and testing \sysname.
We also discuss the challenges we faced during this process.

\subsection{Development process}
\label{sec-lessons-dev}

\vheading{Designing \sysname.}
Our initial design closely followed that of BSD FFS \cite{McKusick99}, but most aspects eventually diverged due to differences between storage hardware and typestate considerations.
We found that some data structures and crash-consistency properties were better suited for use with the typestate pattern than others. 
For example, we chose \sysname's backpointer-based page management approach because it simplifies update dependency rules when allocating or deallocating pages. 
With backpointers, these operations involve a constant number of persistent updates and involve no additional durable structures. 
In contrast, tree or log-based approaches need extra persistent metadata and may require additional updates to balance the tree or free log space, both of which complicate dependencies and typestate management.

An important design decision we had to make was how granular typestate would be. 
One option was to use specific typestates to represent each fine-grained operation; e.g., have one typestate for initializing an inode's link count, another for setting its flags, etc.
Another was to make each typestate more general, with transition functions potentially performing multiple persistent updates.
More general typestates may sacrifice some bug-finding power, but they make the system easier to understand and develop. 
\newcontent{
In \sysname, we aimed to strike a balance by representing only operations that require a specific ordering with typestate. 
For example, when initializing an inode in \sysname, the order in which the values of most fields are set is not relevant to crash consistency, as the contents of the inode are not visible until it is linked into the file system tree.
Therefore, \sysname uses only a single typestate ({\tt Init}) to represent inode initialization, and another ({\tt Committed}) to indicate when it has been added to the tree.
}

\vheading{Parallel model and system implementation.}
We developed the Alloy model alongside \sysname.
This created a useful feedback loop in which the model supported the Rust implementation, and questions or changes to the implementation could be quickly reflected and checked in the model.
We used an incremental development process, incorporating feedback from the Rust compiler and the model immediately as we implemented the system.
Many transitions in the model could be translated directly into Rust typestate transitions, making the model a valuable guide for implementing file system operations.
When we made mistakes translating the model into Rust, typestate checking quickly caught these issues.

Alloy also includes a graphical user interface for visualizing traces of operations on the model.
This was useful for both investigating invariant violations and seeing the set of transitions that occur in a given file system operation, which could be translated directly into system call handler implementations.
It also demonstrated locations where multiple updates could safely share a single store fence, which helped us avoid redundant fences.

\subsection{Finding bugs}
\label{sec-bugs}

While developing \sysname, we used a combination of typestate checking, model checking in Alloy, and dynamic testing to find bugs. 

\vheading{Typestate checking.} 
Typestate checking in the implementation was successful at quickly catching both missing persistence primitives and higher-level ordering bugs; we provide an example of each.

\begin{itemize}[itemsep=0mm]
\item \emph{Missing persistence primitives}.
  Our initial implementation of {\tt write} was missing flush and fence calls after setting the backpointer of a newly-allocated page.
  This bug was immediately highlighted as an error by the compiler.
  Had this bug made it into the implementation, a crash could cause a file to have a size larger than the number of pages associated with it, causing errors when trying to read the file. 

\item \emph{Incorrect ordering}.
  Our initial {\tt rename} implementation mistakenly decremented an inode's link count before clearing the corresponding directory entry.
  A crash could result in a link count that is lower than the true number of links, leading to a dangling link if the inode is subsequently deleted.
\end{itemize}

Although we did not specifically check execution paths without crashes, the crash-consistency invariants encoded in typestate were general enough to
detect some bugs in this code.
For example, the compiler caught a bug where pages were not fully deallocated during {\tt unlink}, which did not require a crash to manifest.
Typestate-related compiler errors were relatively uncommon overall, since using the model as a guide for implementation helped us get ordering right early.
However, it provided a crucial safety net to prevent subtle bugs when we did make mistakes.

\vheading{Model checking with Alloy.}
The Alloy model found several high-level issues in \sysname's design that would have otherwise been difficult to detect and time-consuming to fix, including the following examples. 
\begin{itemize}[itemsep=0mm]
\item We initially believed that crash recovery would not be needed other than to fix \edited{space} leaks. 
  Alloy found an instance of the model where a crash during rename followed by deallocation of the destination directory entry could cause an invalid directory entry to reappear.
  Fixing this required the addition of recovery transitions.
    \item Early designs for \sysname stored . and .. directory entries durably. We discovered via model checking that \edited{our original} dependency rules for handling these directory entries during more complicated operations like {\tt rename} were not correct. Ultimately, we decided to not store these entries, since they can be constructed at runtime using indexed information.
\end{itemize}

\vheading{Testing.}
Neither the typestate pattern nor the Alloy model eliminated the need to test \sysname.
Our primary goal was to check crash-consistency, and we did not check any invariants that only impact regular, non-crash execution.
We used handwritten tests and the xfstests suite \cite{xfstests} to test these unchecked parts of the code.

All bugs found through testing were in parts of \sysname that were not checked by typestate or directly modeled in Alloy.
Most bugs were related to updating volatile indexes or VFS inodes, e.g., failing to remove a deallocated object from an index or setting the wrong value in the VFS inode.
There were also bugs in the implementations of typestate transitions, which are not themselves verified; for example, the transition that wrote new file data to a page did not always calculate the offset for non-aligned writes correctly. 
Implementing bug fixes was quick since we did not need to modify the typestate-restricted interface to objects and there were no proofs to update.

\subsection{Challenges encountered}
\label{sec-challenges}

\vheading{Challenges with typestate.}
\newcontent{
It is easier to write typestate-checked code than it is to write verified code, but this comes at the cost of less powerful compile-time checking.
For example, checking universally-quantified formulas (e.g., all pages in \edited{a file} are allocated) is undecidable, and unlike verification-aware languages, the Rust compiler has no heuristics to attempt to solve them. 
As a result, we cannot ensure invariants such as ``all objects in a set are in a certain typestate'';
specifically, we can't encode this in typestate because the number of objects in the set is not known at compile time.

This became a problem when implementing file-system operations like {\tt unlink}, where we would like to e.g., check that the backpointers of all pages belonging to the file are cleared before deallocating the inode.
Such a check ensures that the system always follows soft updates rule 2 (never re-use a resource before nullifying all previous pointers to it); by clearing all of the page backpointers before deleting the inode, we ensure that none remain when the inode is eventually reused.
However, it is impossible to check this property on arbitrary sets of pages if each page has its own typestate.
We experimented with several workarounds, including forcing {\tt write} operations to update no more than one page at a time (which was prohibitively slow and did not solve the problem for {\tt unlink}), and storing typestate in page structures at runtime and manually adding assertions (which also impacted performance and lost the benefit of static checking). 
Ultimately, we decided to use a single piece of typestate to represent \emph{ranges} of pages (e.g., all of the pages in a file or a contiguous subsection).
\edited{Each typestate operation on such a range performs the operation on all pages in the range.}
This moved some logic into the typestate transitions, making the transition functions themselves more complicated but making page-management logic more centralized and easier to manually audit.
}

\vheading{Challenges with Alloy.}
As \sysname grew in complexity, it became harder to maintain the model and get useful feedback quickly. 
The model checker uses a SAT solver to check invariants, and the formulas representing a large model can take days or weeks to solve.
We checked that traces with multiple concurrent operations were crash consistent, which increased the size of the problem further.
To get faster feedback, we built a custom utility to run multiple independent \edited{instances} of the model checker in parallel and split larger predicates into smaller, more concrete sub-checks. 

It could also be difficult to determine whether a reported failure was a false positive.
A particular challenge was dealing with \emph{frame conditions}, predicates that specify what should not change in a given transition.
Alloy is free to arbitrarily change any state that the current transition does not explicitly mention, so frame conditions are crucial to constrain the model to real traces.
This behavior helps Alloy find corner-case bugs, but it also leads to false positives.
To overcome this challenge, we built a syntax-based checker that parses the model using Alloy's API and checks that each transition explicitly mentions all mutable structures in the model.
The current version of the checker cannot detect all issues, but it detected many missing conditions that would have otherwise taken hours to catch via model checking.

\newcontent{
\subsection{Typestate beyond \sysname}
\vheading{Costs and benefits of typestate.}
We do not have equivalent verified or unverified systems to compare with \sysname in terms of development and debugging effort; however, in the authors' experience, designing and implementing \sysname required more effort than a typical unverified system, but far less effort than a verified storage system.\footnote{\newcontent{For example, author LeBlanc recently worked on a durable log implemented in a verification-aware programming language, which took about 3 months of full-time work.}}
We believe that debugging \sysname was faster and easier than debugging an equivalent unverified system, as following the typestate-enforced ordering rules made it easier to implement the system correctly in the first place and reduced the number of bugs overall. 

Using the typestate pattern for crash consistency represents a useful new point in the tradeoff space between runtime testing and full verification. 
While it comes at the cost of additional development effort compared to unverified systems to determine correct ordering rules and does not gain the same correctness guarantees as verified systems, it does eliminate an entire class of crash consistency bugs that are otherwise difficult to find and fix~\cite{Mohan18, Kim2019, LeBlanc23}.
Furthermore, as the pattern builds ordering rules directly into a system's implementation, the rules will stay up to date and continue to prevent crash-consistency bugs as the system is developed further~\cite{hawblitzel2015, Newcombe2015}. 

\vheading{Broader applicablity.} 
As the typestate pattern is a general approach for statically checking the order of updates to data structures, it is useful in a broad variety of contexts, several of which are described below.
\begin{itemize}
    \item Volatile data structures: \sysname does not use typestate to manage updates to volatile data structures, but prior work on typestate verification has focused entirely on such use cases~\cite{Strom86, aldrich2009}.
    \item Other types of storage systems: The typestate pattern could be used to enforce ordering invariants on durable updates in other types of storage systems (e.g., key-value stores) with different crash-consistency mechanisms. We note that crash-consistency mechanisms like journaling and copy-on-write do not achieve consistency entirely through ordering and would require auxiliary techniques to check properties like atomicity.
    \item Durable layout: \sysname's on-storage layout is tailored to reduce the number of durable updates per file-system operation and to simplify ordering rules. Other layouts could also be used in typestate-checked storage systems, although the complexity of the ordering rules would increase.
    \item Asynchrony: The typestate pattern is compatible with asynchronous systems, although the ordering rules to enforce are much more complicated in such systems, as updates from different operations may be interleaved.
\end{itemize}
}

%% file: sections/evaluation/evaluation4.tex
\section{Evaluation}
\label{sec:eval}

We seek to answer the following questions in our evaluation of \sysname:

\begin{enumerate}[itemsep=0mm]
\item What is the latency of different file-system operations on \sysname? (\sref{sec:microbench})
    \item How does \sysname perform on macrobenchmarks? (\sref{sec:macrobench})
    \item How does \sysname perform on real applications? (\sref{sec:applications})
    \item How long does \sysname take to mount and recover from crashes? (\sref{sec:mount_time})
    \item What compilation, memory, and CPU overheads does \sysname incur? (\sref{sec:overhead})
    \item Is \sysname correct? (\sref{sec:correctness})
\end{enumerate}

\subsection{Experimental setup}
\label{sec:setup}

We evaluate \sysname on a two-socket, 32 core machine with 128GB of memory and one 128GB Intel Optane DC Persistent Memory Module. 
The evaluation machine runs Debian Bookworm and Linux 6.3.

We compare \sysname against ext4-DAX \cite{daxfiles}, NOVA \cite{Xu16}, and WineFS \cite{Kadekodi21}. 
We configure all three systems to provide metadata consistency but not data consistency to match \sysname's guarantees. 
We cannot compare \sysname to SoupFS \cite{Dong17}, the only other soft updates PM file system, as it is not open source. 
Due to time constraints, we were unable to compare against the recent ArckFS~\cite{Zhou23} file system. We hope to do so in the future.
\edited{All reported results are the average of multiple trials. The red errors bars in Figure~\ref{fig:all_graphs} indicate the minimum and maximum values recorded over all trials.}

\input{sections/figures/all_graphs}

\subsection{Microbenchmarks}
\label{sec:microbench}

We compare each system's latency by testing several file system operations: appending and reading 1KB and 16KB to a file, file creation, directory creation, renaming a directory, and unlinking a 16KB file. 
None of the tests call {\tt fsync}.

The average latency \edited{over 10 trials} of the tested operations are shown in Figure~\ref{fig:all_graphs}(a).
The lowest latency file system in each test is either WineFS or \sysname. 
Ext4-DAX has the highest latency on many operations because it interacts with the Linux kernel block layer for tasks like block allocation\edited{, which incurs additional software overhead.}
It achieves similar performance to the other systems on operations that do not go through the block layer (e.g., unlink).
NOVA has higher latency on {\tt mkdir} and {\tt rename} than WineFS and SquirrelFS because operations that update multiple inodes require journaling in NOVA. 

\subsection{Macrobenchmarks}
\label{sec:macrobench}

We evaluate \sysname on the Filebench \cite{filebench} storage benchmark suite.
We run four workloads from the suite -- fileserver, varmail, webserver, and webproxy -- in their default configurations.
Fileserver performs mostly writes with some whole file reads; varmail is half appends and half reads; webproxy appends to each file and reads from it several times; and webserver reads and occasionally appends to a log file.
Figure~\ref{fig:all_graphs}(b) shows the average throughput in kops/sec for each file system on each workload. 
\sysname achieves slightly better throughput than the next fastest system on fileserver and varmail (8\% and 13\% better, respectively) and within 10\% of the fastest system on both webserver and webproxy.
Fileserver and varmail perform many small appends, which \sysname performs well on due to its lack of journaling.
Webserver and webproxy are more read-heavy, which all systems perform roughly equally on.
Ext4-DAX does not go through the block layer on reads and it benefits from data contiguity awareness, making its performance similar or better than the other systems on these workloads.

\subsection{Applications}
\label{sec:applications}
 
\vheading{YCSB on RocksDB.} We evaluate the four systems on RocksDB \cite{rocksdb} using YCSB workloads \cite{ycsb}. 
We run all workloads on a 25GB database with 25M records, 25M operations, and 8 threads.
All workloads are run using standard workload configurations and the default settings of YCSB, which uses system calls for all operations.
Figure~\ref{fig:all_graphs}(c) shows throughput in kops/second relative to Ext4-DAX on each tested workload.

\sysname outperforms the other systems on Loads A and E, which are 100\% small inserts. 
As seen on the other benchmarks, \sysname performs particularly well on small appends due to its lack of journaling or logging.
Writes that require page allocation are particularly expensive in the other systems, as journaling/logging the new metadata incurs an additional 2-3us in NOVA and WineFS and 3-4us in Ext4-DAX.
Ext4-DAX and NOVA both also journal or log metadata on every append, spending roughly 30\% of each non-allocating call (approx 1-1.5us) managing journals/logs.

All file systems are within 10\% of Ext4-DAX's throughput on Runs B, C, and D. All of these workloads are at least 95\% small (4KB) reads, which all four systems achieve similar performance on.

\sysname achieves the best throughput on Runs A and F, which are 50\% reads and 50\% updates (Run A) or read-modify-write operations (Run F).
Ext4-DAX, NOVA, and WineFS all incur logging/journaling on these workloads; Ext4-DAX outperforms NOVA and WineFS because it has less journaling overhead for in-place updates and is more aware of data contiguity on reads.

Ext4-DAX achieves the best performance on Run E, which is 95\% range scans and 5\% inserts.
Ext4-DAX's contiguity-awareness and better fragmentation-prevention mechanisms help it outperform the other systems on larger read operations.

\vheading{LMDB.} We also run LMDB~\cite{LMDB}, a memory-mapped database, using {\tt db\_bench}'s {\tt fillseqbatch}, {\tt fillrandbatch}, and {\tt fillrand} workloads.
Each experiment uses 100M keys on an empty file system.
Figure~\ref{fig:all_graphs}(d) shows the throughput in kops/sec for each file system on each workload.
Each file system has throughput with 12\% of the other systems.
Most updates are done to memory-mapped files, so differences in the performance of system calls and metadata management designs have a reduced impact.

\vheading{Git.} We also evaluate the performance of \sysname by performing {\tt git checkout} of major Linux kernel versions.
The time to check out a given version in each file system is within 8\% of the other systems. 

\subsection{Mount time}
\label{sec:mount_time}

\sysname takes longer to mount than other PM file systems because it must rebuild volatile indices for the entire file system.
Table~\ref{tab:mount_time} shows how long it takes to mount \sysname on a 128GB PM device with different contents. 
The $\approx5.5$ seconds it takes to initialize or mount an empty system is the overhead of zeroing or scanning the metadata tables and creating volatile allocators.
We also measure the time to mount a system with 100\% data and inode utilization.
Most of this time is spent allocating space for and managing the volatile indexes and allocators.

If \sysname detects that it was not unmounted cleanly, it constructs additional structures to keep track of orphaned objects and the true link count of each inode.
It fills in these structures during the regular rebuild scan and uses them to free orphans and correct link counts at the end of the mount process.
\sysname also checks each directory entry for non-null rename pointers and either rolls back or completes any interrupted renames.
Table~\ref{tab:mount_time} reports the time it takes \sysname to perform recovery scans on a cleanly-unmounted device. 
Mounting with recovery takes longer than a standard mount because the file system must construct orphan-tracking structures and do an extra iteration over all directories to check for rename pointers in addition to building the volatile indices and allocators.

\sysname's mount time could be improved by parallelizing some of its rebuild and recovery logic. 
For example, the inode and page descriptor table scans are completely independent and could be done in parallel. 
The file system tree rebuild logic could also be distributed across multiple threads.

\input{sections/figures/mount_time}

\subsection{Resource usage}
\label{sec:overhead}

\vheading{Compilation.}
\sysname takes approximately 10 seconds to compile on our test machine, including typestate checking. This compares well to fully-verified systems; FSCQ \cite{Chen15} takes about 11 hours to verify, and VeriBetrKV \cite{Hance20} takes 1.8 hours (10 minutes when parallelized). 

\sysname also compiles faster than the other tested systems on the test machine.
Table~\ref{tab:compilation_time} shows the size of each system in lines of code and how long it takes to compile.
\sysname's more complicated typechecking does not noticeably impact its compilation time.

\begin{table}[]
    \centering
    \begin{tabular}{l c c}
        \toprule
        System & LOC & Compile time (s)  \\
        \midrule
        Ext4 & 45K & 38 \\
        NOVA & 16K & 20 \\
        WineFS & 9K & 13 \\ 
        \sysname & 7.5K & 10 \\ 
        \bottomrule
    \end{tabular}
    \caption{Time to compile different PM file systems as loadable kernel modules. Ext4's line count includes interleaved DAX and non-DAX code.}
    \label{tab:compilation_time}
\end{table}

\vheading{Memory.} \sysname maintains indexes for fast lookups of files and directory entries.
Each regular file has an index mapping its inode number (8 bytes) to each of its pages and their offsets (16 bytes total).
Thus, the index entries for a 1MB file use about 4KB of memory.
Each directory has a similar inode to page index (without offsets), \edited{plus a mapping} from directory entry names to metadata like their location on PM and inode number.
The current maximum name length is 110 bytes (which makes directory entries 128 bytes) and \sysname does not currently hash or compress names.
Therefore, each directory entry takes up approximately 250 bytes in the index.

\vheading{CPU.} \sysname does not start new threads in any of its operations. We leave the use of more threads for operations like freeing pages, running crash recovery, etc. to future work.

\subsection{Correctness}
\label{sec:correctness}

\vheading{Model checking.} We check that a correctness invariant always holds in all traces of our Alloy model. 
We bound traces to include two operations (which may be concurrent), 10 persistent objects, and up to 30 steps. 
The invariant includes both sanity checks on the model as well as file system consistency checks. The sanity checks ensure, for example, that objects will never end up with conflicting typestates.
The consistency checks ensure that 1) objects always have a legal link count, 2) there are no pointers to uninitialized objects, 3) freed objects do not contain pointers to other objects, and 4) there are no cycles of rename pointers and directory entries are pointed to by at most one rename pointer.

\vheading{Testing.} We test \sysname using a set of handwritten tests and the xfstests~\cite{xfstests} test suite. 
\newcontent{\sysname currently passes all supported tests \edited{(67)} from xfstests' {\tt generic} test suite.}
\edited{The rest of the tests use system calls or arguments that are currently not supported by \sysname.}

\vheading{Crash consistency.}
\newcontent{
We used Chipmunk~\cite{LeBlanc23} to test \sysname for crash-consistency bugs. 
We modified Chipmunk's test generators to remove several system calls that \sysname does not currently support but otherwise ran its full suite of systematically-generated tests and fuzzed the system for approximately 24 hours. 
Chipmunk did not find any ordering-related crash-consistency bugs in \sysname, providing evidence that typestate-checked SSU is an effective mechanism for preventing such bugs. 
Chipmunk did find four crash consistency bugs in unchecked parts of \sysname code, three in its rebuilding of volatile data structures and one in the \edited{body of typestate transitions} in which a cache line flush was issued to the wrong address.
As these are not caused by incorrect update ordering, the typestate pattern did not catch them at compile time.
We found that using the typestate pattern in \sysname made locating and fixing these bugs faster and easier, as we could focus on the specific regions of code that are unchecked and are thus more likely to have bugs.
}

\subsection{Summary}

\sysname provides comparable performance to other PM file systems, while providing strong guarantees about its crash consistency.
Due to the innovative use of typestate checking, we were able to implement SSU and gain confidence in its correctness.
\sysname gains an advantage over other file systems in write-dominated workloads, since soft updates avoids writing to a log or to a second copy of the data.
The design of \sysname trades off good common-case performance for slightly longer mount times compared to other file systems; we believe this is acceptable since crashes are rare.
\sysname compiles at the same rate as other PM file systems, despite the strong type checking.

%% file: sections/figures/all_graphs.tex
\begin{figure*}[t]
    \centering
    \includegraphics[width=0.9\textwidth]{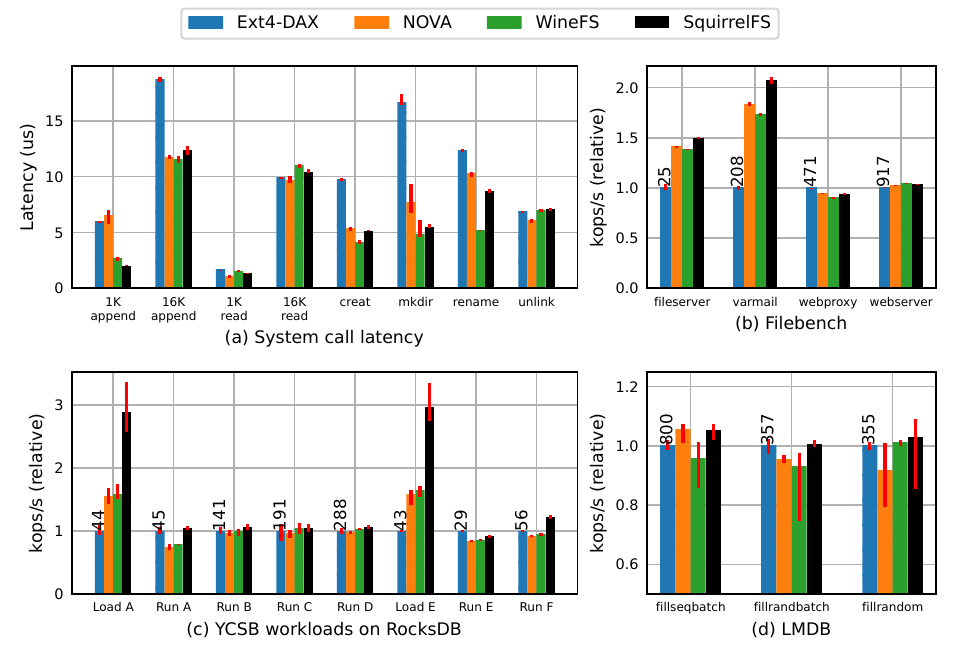}
    \caption{This figure shows the performance of the evaluated file systems on different benchmarks and applications. (a) shows absolute latency of different file system operations; (b), (c), and (d) show the relative throughput in kops/s of each system relative to Ext4-DAX on filebench, YCSB on RocksDB, and LMDB respectively.}
    \label{fig:all_graphs}
\end{figure*}

%% file: sections/figures/mount_time.tex
\begin{table}
\centering 
\begin{tabular} {c | l | c}
    \toprule 
                                    & System state  & Mount time (s) \\ \midrule 
    \multirow{3}{4em}{Normal mount} & mkfs          & 5.80 \\
                                    & Empty         & 5.51 \\
                                    & Full          & 30.50 \\
    \midrule
    \multirow{2}{4em}{Recovery mount}   & Empty     & 5.76 \\
                                        & Full      & 55.50 \\
    \bottomrule
\end{tabular}
\caption{Time in seconds to mount \sysname file system images in differrent states. Times in the recovery mount column come from mounting a cleanly-unmounted file system that runs a recovery scan in addition to normal rebuild scans.}
\label{tab:mount_time}
\end{table}

%% file: sections/related/related3.tex
\section{Related work}
\label{sec:related}

\vheading{Rust for PM}.
\sysname was inspired by Corundum~\cite{Hoseinzadeh21}.
Corundum builds data structures whose low-level properties are checked using Rust's type system. 
For example, Corundum ensures that there are no pointers to volatile memory stored in persistent memory,
and that persistent state is only updated in transactions.
It focuses on lower-level persistent memory programming errors and cannot prevent higher-level logical bugs. 
Corundum also requires \emph{all} updates to PM to be in transactions, which is overly restrictive for many systems.
In contrast to Corundum, \sysname checks high-level file-system crash-consistency properties using type-checking without placing constraints on how the file system is used.

\vheading{Soft updates for PM}.
Two PM file systems use soft updates for crash consistency: SoupFS~\cite{Dong17} and ArckFS~\cite{Zhou23}.
Unlike \sysname, SoupFS is asynchronous and uses background threads to flush updates.
It uses byte-addressable updates to eliminate cyclic dependencies.
ArckFS is a user-space PM file system built on the Trio architecture that uses synchronous, soft-updates-esque updates for simple operations (e.g., creating a file) and undo journaling in more complicated cases.
Unlike ArckFS, \sysname uses only synchronous soft updates for its crash consistency;
the novel way in which \sysname implements atomic rename (without journaling or copy-on-write) further differentiates it from ArckFS.
Both SoupFS and ArckFS are written in C, and do not use Rust's type system to check their crash consistency.

\vheading{Storage systems in Rust}.
Bento~\cite{Miller21} is a framework for building in-kernel file systems in Rust.
The corresponding file system from the Bento project, BentoFS, was designed for block devices.
Bento does not utilize the type system of Rust to check file-system properties.

ShardStore~\cite{bornholt:shardstore} is a Rust key-value store used in Amazon S3 that uses an asynchronous soft-updates-inspired crash-consistency mechanism. 
The rules for when something should be written to storage in ShardStore were checked with DepSynth~\cite{VanGeffen23}, a tool for synthesizing soft updates dependency rules.
Unlike ShardStore, \sysname uses a synchronous version of soft updates, and provides higher-level primitives like atomic rename; ShardStore does not utilize the type system to perform higher-level checks.

%% file: sections/conclusion1.tex
\section{Conclusion}
\label{sec:conclusion}

This paper presents a new methodology for crash-consistent file system development.
We propose the use of the typestate pattern in Rust to statically check crash-consistency invariants with low proof burden.
We also introduce a novel crash-consistency mechanism, synchronous soft updates, that is well-suited to enforcement with the typestate pattern and that
eliminates many challenges associated with the original soft updates technique. 
We develop \sysname, a new file system for persistent memory that uses statically-checked synchronous soft updates for crash consistency.
\sysname achieves comparable or better performance than other PM file systems and required no language modifications or verification expertise to build.
\sysname, its Alloy model, and our Alloy utilities are available at \url{https://github.com/utsaslab/squirrelfs}.